# Optically controlled spin-polarization memory effect on Mn delta-doped heterostructures

M. A. G. Balanta [1,2,*], M. J. S. P. Brasil[1], F. Iikawa[1], Udson C. Mendes [1,3],
J. A. Brum[1], Yu. A. Danilov[4], M. V. Dorokhin[4], O. V. Vikhrova[4], and B. N. Zvonkov[4]

[1] *Instituto de Física "Gleb Wataghin", Unicamp, 13083-859 Campinas, SP, Brazil*
[2] Departamento de Física, Universidade Federal de São Carlos, CP 676, São Carlos, SP 13565-905, Brazil
[3]Laboratoire Pierre Aigrain, École Normale Supérieure-PSL Research University,CNRS, Université Pierre et Marie Curie-Sorbonne Universités,Université Paris Diderot-Sorbonne Paris Cité, 24 rue Lhomond, 75231 Paris Cedex 05, France.
[4] *Research Institute, State University Nizhny Novgorod, Russia*

*Email: magbfisc@ifi.unicamp.br

## ABSTRACT

We investigated the dynamics of the interaction between spin-polarized photo-created carriers and Mn ions on InGaAs/GaAs: Mn structures. The carriers are confined in an InGaAs quantum well and the Mn ions come from a Mn delta-layer grown at the GaAs barrier close to the well. Even though the carriers and the Mn ions are spatially separated, the interaction between them is demonstrated by time-resolved spin-polarized photoluminescence measurements. Using a pre-pulse laser excitation with an opposite circular-polarization clearly reduces the polarization degree of the quantum-well emission for samples where a strong magnetic interaction is observed. The results demonstrate that the Mn ions act as a spin-memory that can be optically controlled by the polarization of the photocreated carriers. On the other hand, the spin-polarized Mn ions also affect the spin-polarization of the subsequently created carriers as observed by their spin relaxation time. These effects fade away with increasing time delays between the pulses as well as with increasing temperatures.

# INTRODUCTION

GaMnAs alloys, where the Mn dopant supply both a magnetic moment and a spin-polarized carrier, have attracted considerable interest as a spintronics material [1, 2]. Writing and optical readout of the Mn spin in diluted magnetic semiconductors is a major point for practical applications [3-5]. Much attention has been devoted to the optical orientation of the Mn acceptors and their spin manipulation in semiconductors [6,7]. This effect has been investigated mainly in GaAs: Mn bulk samples by resonant excitation of electrons from residual donors in the vicinity of a Mn acceptor or in quantum dots, where the coupling between the carriers and the Mn ion is reinforced by their strong overlap [8, 9].

Quantum well (QW) structures with a Mn delta-doped ($\delta_{Mn}$) layer at the barrier were proposed as a solution to preserve the optical properties of the QW confined carriers without destroying their interaction with the magnetic ions [10,11]. In spite of the reduced overlap, it was demonstrated that the spin interaction survives on a structure consisting of an InGaAs/GaAs QW with a $\delta_{Mn}$ at the GaAs barrier [12-14]. In this work, we investigate the time dynamics of this interaction in a series of samples. We developed a special technique involving two pulsed beams with individually controlled circular-polarizations and a variable time-delay between them. We observed that the application of a pre-pulse with an opposite polarization gives rise to an asymmetry of the polarization degree of the QW emission. The results indicate that the spin-polarized carriers created by the first pulse affect the spin-polarization of the Mn ions, which in turn affects the spin-polarization of the carriers generated by the second pulse. It is thus possible to optically control the spin of Mn ions through the polarization of carriers photogenerated in a nearby QW and use them as a spin-polarization memory.

# EXPERIMENTALS DETAILS

The experiments were performed in structures consisting of an InGaAs QW with a $\delta_{Mn}$ layer at the GaAs barrier as shown in Fig. 1. We investigated two set of samples that differ by the inclusion of a C delta-doping layer ($\delta_C$) at the other QW barrier, which provides additional holes to the InGaAs QW. In the first set, without C, the Mn doping position was varied, while in the second set, with C, the Mn concentration was varied. In the following, we refer to the first and second set of samples as MN- and CMN- series, respectively. All samples were grown using a hybrid system combining metal-organic chemical vapor and pulsed-laser ablation depositions. First, an undoped GaAs buffer layer, the $In_{0.16}Ga_{0.84}As$ QW (10 nm) and a

GaAs spacer layer ($d_S$) were grown by MOCVD at a high temperatures (~600 °C). The precursors were trimethylgallium, trimethylindium and arsine. On the CMN- series, carbon tetrachloride doping was used to grow the $\delta_C$ separated by a 10 nm GaAs layer from the InGaAs QW. On the second stage, we have used a Q-switched YAG: Nd laser ablation system with Mn and GaAs targets at temperatures $T_{Mn}$ for growing the Mn delta-doping layer and the GaAs capping layer ($d_C$), respectively. All the growth was performed in the same reactor. Further details of the growth can be found in Ref. [13]. A complete list of the growth parameters from the investigated samples is presented in Table 1.

Time-resolved photoluminescence (PL) measurements were performed using a fs Ti:Sa laser and a streak-camera system (time resolution ~50 ps). The laser wavelength was tuned for resonant QW excitation. The right- ($\sigma^+$) and left- ($\sigma^-$) circular-polarized components of the excitation beams and the optical emission were selected with appropriated optics. The circular polarization of each beam can be selected independently. The time delay $\Delta t$ between the pulses from the two beams was controlled by changing the optical path of one of the beams. From now on we refer to the pulses of the beam that arrive a time $\Delta t$ before the pulses of the other beam as the pre-pulses. The results presented here correspond to the condition where the pre-pulses are $\sigma^-$ polarized, and the following pulses from the second beam are $\sigma^+$ polarized. Measurements with opposite polarizations were also performed and gave equivalent results.

The degree of polarization of the PL emission is defined as:

$$Pol = (I^{\sigma+} - I^{\sigma-}) / (I^{\sigma+} + I^{\sigma-}) \qquad (1)$$

where $I^{\sigma+/\sigma-}$ is the intensity of the $\sigma^{+/-}$ emission component.

## RESULTS AND DISCUSSION

Table 1 presents the PL decay time ($\tau$) and the spin-relaxation time ($\tau_s$) obtained by fitting the time-resolved PL transients using only one excitation beam by simple exponential decays. We point out that all samples present similar values of $\tau$ of order of 0.2 ns, which is consistent with the results obtained for similar QWs [15] and somehow larger than our experimental resolution. Therefore we are not very confident on doing a detailed analysis based on the small variations observed for $\tau$. However, $\tau_s$ ranges from less than 1 ns up to more than 2 ns, which we will discuss below.

Figure 2 shows typical time-resolved measurements from sample MN1 (see Table 1) using two excitation beams. The excitation energy is ~30 meV above than the PL peak (1.383 eV) with an averaged power of 10 mW. We point out that during the laser pulse the excitation power is ~$10^4$ stronger than the measured averaged power as estimated considering the ratio between the pulse duration (~100 fs) and the laser repetition (~12 ns). The experiment was performed using a time delay of $\Delta t$=500 ps between the two excitation beams as indicated by the diagram of Fig. 2. The streak camera images were obtained by measuring the $\sigma^+$ and $\sigma^-$ circularly-polarized components of the PL emission. The aim of our experiment is to use the pre-pulse to induce a particular spin-polarization of the Mn ions, and to use the following pulse after a time $\Delta t$, to probe the effects of the Mn polarization on the polarization degree of the PL emission from the QW.

Figures 3a and 3b show the transients of $\sigma^-$ and $\sigma^+$ components of the integrated QW PL emission from sample MN1 using one and two excitation beams. With a single $\sigma^+$ excitation beam, the PL transients show an expected dominance of the $\sigma^+$ emission giving rise to an initial polarization degree of *Pol* ~ 50% immediately after the laser pulse, as shown in Figure 3c. When the second excitation beam is turned on giving rise to $\sigma-$ pre-pulses with $\Delta t$=500 ps, the polarization of the PL emission is initially negative, around -50 %, as shown in Fig. 3c. Thus, when the $\sigma^+$ excitation pulse hits the structure there is a residual PL emission with $\sigma^-$ dominance. The $\sigma^+$ excitation pulse inverts this state, so that the PL polarization degree *Pol* changes from negative to positive.

Figure 3c reveals a rather interesting effect. As pointed out, the magnitude of the polarization degree immediately after a laser pulse for a single beam excitation condition is ~50% for sample MN1. However, when the pre-pulse is applied, the polarization degree immediately after the $\sigma^+$ excitation pulse becomes surprisingly smaller (~ 25%). In fact, a correction to take into account the presence of the residual carriers from the pre-pulse is necessary. The corrected-polarization degree is also shown in Fig. 3c, which was calculated by subtracting the estimated PL intensity created by the pre-pulse considering a mono-exponential decay (Figure 3b) from the PL intensity measured after the second pulse. As shown in Figure 3c, by doing this correction the initial polarization degree immediately after the $\sigma^+$ excitation pulse becomes ~30%, which is still significantly smaller than the ~50% value from the measurements without a pre-pulse.

We define here a parameter, $\Delta Pol$, as the difference between the modulus of the initial polarization degree when the sample is excited with a single excitation beam ($Pol_0$) and the modulus of the initial polarization degree under the presence of a pre-pulse, considering the correction discussed above concerning the subtraction of the residual PL intensity generated by the pre-pulse. This parameter is directly shown in Figure 3c for the sample MN1. As the value of $Pol_0$ can vary from sample to sample, we will analyze here the relative effect of the pre-pulse through the ratio ($\Delta Pol / Pol_0$). This measured ratio as a function of the time separation $\Delta t$ is presented in Figure 4 for all investigated samples. We observe two main results. First, the effect of the pre-pulse decreases with $\Delta t$ for all samples. Second, we notice that the two series of samples show a rather distinct behavior. The effect of the pre-pulse on the CMN samples is significantly smaller as compared to the samples from the MN samples. A reference sample, without Mn, shows no effect at all, as expected.

We believe that these results are an indication of an optical control of the Mn spin polarization. In this interpretation, the spin-down polarized carriers created by the $\sigma^-$ pre-pulse interact with the Mn ions giving rise to an effective spin-down polarization of the magnetic ions. Conversely, the polarization of the Mn ions affects the spin-polarization of the carriers created by the following $\sigma^+$ pulse. This fast process occurs during the rising of the PL emission so that it cannot be resolved by our experimental set-up, but it gives rise to a reduced initial polarization degree associated to the pre-pulse. As the Mn ions should present relatively long spin times compared to the spin of the carriers, they must act as an effective spin reservoir. As the delay time between the pulses is increased, the Mn ions lose its spin polarization, becoming less effective in reversing the spins of the photocreated carriers, which explains the decreasing of ($\Delta Pol/Pol_0$) with increasing $\Delta t$ as shown in Figure 4. A quantitative analysis of the complex relation between ($\Delta Pol/Pol_0$) and the effective polarization of the Mn ion is prevented by the fact that we do not have access to the polarization dynamics during the laser pulse. However, the non-zero values of ($\Delta Pol/Pol_0$) obtained for $\Delta t=1.5$ ns indicate that the Mn spin lifetimes are longer than 1 ns in our structures. We point out that we have also performed measurements using the same circular polarization degree for both the pulse and the pre-pulse, not shown here. We observed that under this condition ($\Delta Pol/Pol_0$) becomes essentially zero. This excludes heating effects caused by the pre-pulse.

The polarization effect is stronger for sample MN1 as compared to sample MN2, which is consistent with the larger separation between the $\delta_{Mn}$ layer and the QW for MN2. This result supports our interpretation that the observed effect originates from the spin coupling between Mn ions and confined carriers, so that it is reduced when the overlap between these entities decreases. We also remark that ($\Delta Pol/Pol_0$) is significantly smaller for the CMN samples as compared to the MN samples and is null for the reference sample without Mn. We interpret the reduced effect on the CMN series as an indication of a reduced interaction between confined carriers and Mn ions on these samples. This conclusion is supported by previous independent results that also indicated a reduced overlap on those samples [10,13]. In addition to provide additional holes to the QW, the C delta-doping layer modifies the self-consistent potential profile of the structure, which changes the wave-function overlap [10]. Besides, CMN samples were grown in a slightly different temperature.

This should affect the Mn distribution and may also contribute to the reduced overlap on these samples.

We also point out that even for measurements with a single excitation beam, the Mn ions should act as a spin reservoir. In this case, the carriers with a well defined spin-polarization created by one excitation pulse should interact with the Mn ions during its transient. This interaction should result in an effective polarization of the Mn ions, which in turn, should contribute to a longer spin polarization time of the photocreated carriers. Thus, samples with stronger interaction between the confined carriers and the Mn ions should present longer spin times, which is indeed observed as we compare the results from Table 1 and Figure 4. This effect explains the relatively larger spin relaxation times ($\tau_s$) obtained for the MN-samples as compared to the CMN samples. The correlation is also consistent when we compare samples MN1 and MN2, as the first one presents the larger values of ($\Delta Pol / Pol_0$) and the longer $\tau_s$.

Finally, we studied the temperature dependence of $\tau_s$ and ($\Delta Pol/Pol_0$) for a constant $\Delta t=0.6$ ns for the sample MN1. Both parameters decrease with increasing temperatures as shown in Figure 5. Albeit reduced, the Mn spin memory effect is observed on temperatures up to 100 K. We do expect that the scattering mechanisms by phonons become more efficient with increasing temperatures, resulting in faster spin-relaxation of the carriers via the Elliot-Yafet mechanism, which is consistent with the reducing spin lifetimes observed in Figure 5. At the same time, ($\Delta Pol/Pol_0$) also decreases with the temperature. Two correlated effects might be attributed to this behavior. On one hand, the decreasing of $\tau_s$ due to scattering mechanisms must reduce the efficiency of the spin-polarized photocreated carriers to orientate the Mn spins, thus reducing the effective magnetization of the Mn ions and the resulting value of ($\Delta Pol/Pol_0$). On the other hand, increasing temperatures should also reduce the magnetization of the Mn ions per se, and therefore, diminishing ($\Delta Pol/Pol_0$). Furthermore, the decrease of the spin polarization time of the Mn ions with temperature can also contribute to the decreasing of $\tau_s$ due to the reduction of the spin reservoir [16].

**SUMMARY**

In conclusion, we observed a clear effect of spin memory for samples where the QW confined carriers present a significant interaction with the Mn ions from a nearby $\delta_{Mn}$ layer. For those samples, we demonstrated that by applying a pre-pulse from an additional excitation beam with an opposite circular polarization gives rise to a reduced polarization degree of the PL emission as compared to the results without the pre-pulse. We propose that the pre-pulse writes the spin information on Mn ions that act as a spin reservoir, and this memory is read by the second pulse as a reduced polarization degree. Furthermore, the Mn spins also act as a spin memory that increases the spin relaxation time of the photocreated carriers on measurements without the pre-pulse. The results demonstrate that despite the relatively large spatial separation between the confined carriers and the Mn ions in our samples, their magnetic interaction persists and gives rise to the possibility to manipulate the spins of the Mn ions optically.

**Acknowledgments**

We acknowledge the Brazilian financial agencies CNPq (Project No. 149365/2010-1, 229659/2013-6), CAPES, FAPESP (Proc. 2011/20985–6, 2011/50975-2 and 2010/11393–5), the Ministry of Education and Science of Russian Federation (Project No 8.1054.2014/K) and the Russian Foundation for Basic Research (Grants No 13-07-00982a, 13-02-97140-reg). The technical support from M. Tanabe is kindly acknowledged.

# REFERENCES

1. Dielt, T & Ohno, H. Dilute ferromagnetic semiconductors: Physics and spintronic structures. *Rev. Mod. Phys.* **86** 187 (2014).

2. Jungwirth, T. et al. Theory of ferromagnetic (III,Mn)V semiconductors. *Rev. Mod. Phys*. **78**, 809 (2006).

3. Kobak, J. et al. Designing quantum dots for solotronics. *Nat. Commun*. **5**, 3191 (2014).

4. Jungwirth, T. et al. Spin-dependent phenomena and device concepts explored in (Ga, Mn) As. *Rev. Mod. Phys*. **86**, 855 (2014).

5. Hai, P. N., Maruo, D. & Tanaka, M. Visible-light electroluminescence in Mn-doped GaAs light-emitting diodes. *Appl. Phys. Lett*. **104**, 122409 (2014).

6. M. Goryca, M. et al. Optical Manipulation of a Single Mn Spin in a CdTe Based Quantum Dot. *Phys. Rev. Lett*., **103**, p. 087401 (2009).

7. Žutić, I., Fabian, J. & Das Sarma, S. Spintronics: Fundamentals and applications. *Rev. Mod. Phys*. **76**, 323 (2004).

8. Besombes, L. et al. Optical control of the spin of a magnetic atom in a semiconductor quantum dot. *Nanophotonics* **4**: 75–89 (2015).

9. Akimov, I. A. et al. Electron spin dynamics and optical orientation of Mn2+ ions in GaAs. *J. Appl. Phys*. **113** 136501 (2013).

10. Mendes, U. C., Balanta, M. A. G., Brasil, M. J. S. P. & Brum, J. A. Electronic and optical properties of InGaAs quantum wells with Mn-delta-doping GaAs barriers. *arXiv preprint arXiv*:1509.07136 (2015).

11. Gazoto, A. L. et al. Enhanced magneto-optical oscillations from two dimensional hole-gases in the presence of Mn ions. *Appl. Phys. Lett*. **98**, 251901 (2011).

12. Korenev, V. L. et al. Dynamic spin polarization by orientation-dependent separation in a ferromagnet–semiconductor hybrid. *Nature Commun*. **3**, 959 (2012).

13. Balanta, M. A. G. et al. Effects of a nearby Mn delta layer on the optical properties of an InGaAs/GaAs quantum well. *J. Appl. Phys*. **116**, 203501 (2014).

14. Rozhansky, I. V. et al. Spin-dependent tunneling in semiconductor heterostructures with a magnetic layer. *Phys. Rev. B*. **92**, 125428 (2015).

15. Balanta, M. A. G. et al. Compensation effect on the CW spin-polarization degree of Mn-based structures. *J. Phys. D: Appl. Phys*. **46**, 215103 (2013).

16. Dorokhin, M. V. et al. The circular polarization inversion in δ⟨Mn⟩/InGaAs/GaAs light-emitting diodes. *Appl. Phys. Lett*. **107**, 042406 (2015).

Table 1. Parameters of investigated samples and measured electron lifetime ($\tau$) and spin-relaxation time constants ($\tau_s$).

|            | Sample | $d_s$,nm | $Q_{Mn}$, ML | $d_c$, nm | $T_{Mn}°C$ | $\tau$ (ps) | $\tau_s$ (ps) |
|---|---|---|---|---|---|---|---|
| MN series  | MN1    | 4 | 0.30 | 12 | 400 | 237 | 2150 |
|            | MN2    | 8 | 0.30 | 12 | 400 | 189 | 1960 |
| CMN series | CMN0   | 3 | 0.00 | 30 | 450 | 180 | 562  |
|            | CMN1   | 3 | 0.13 | 30 | 450 | 190 | 870  |
|            | CMN2   | 3 | 0.20 | 30 | 450 | 120 | 950  |

**FIGURES**

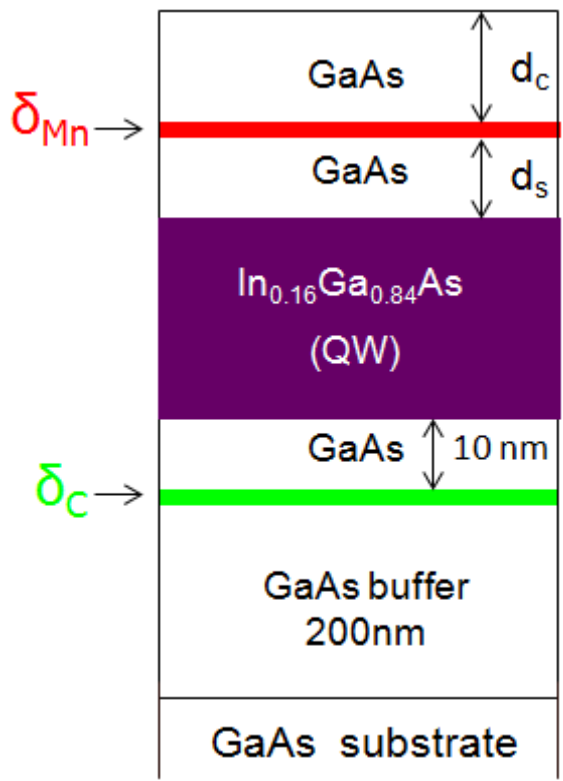

Figure 1. (Color Online) Schematic diagram of the investigated structures.

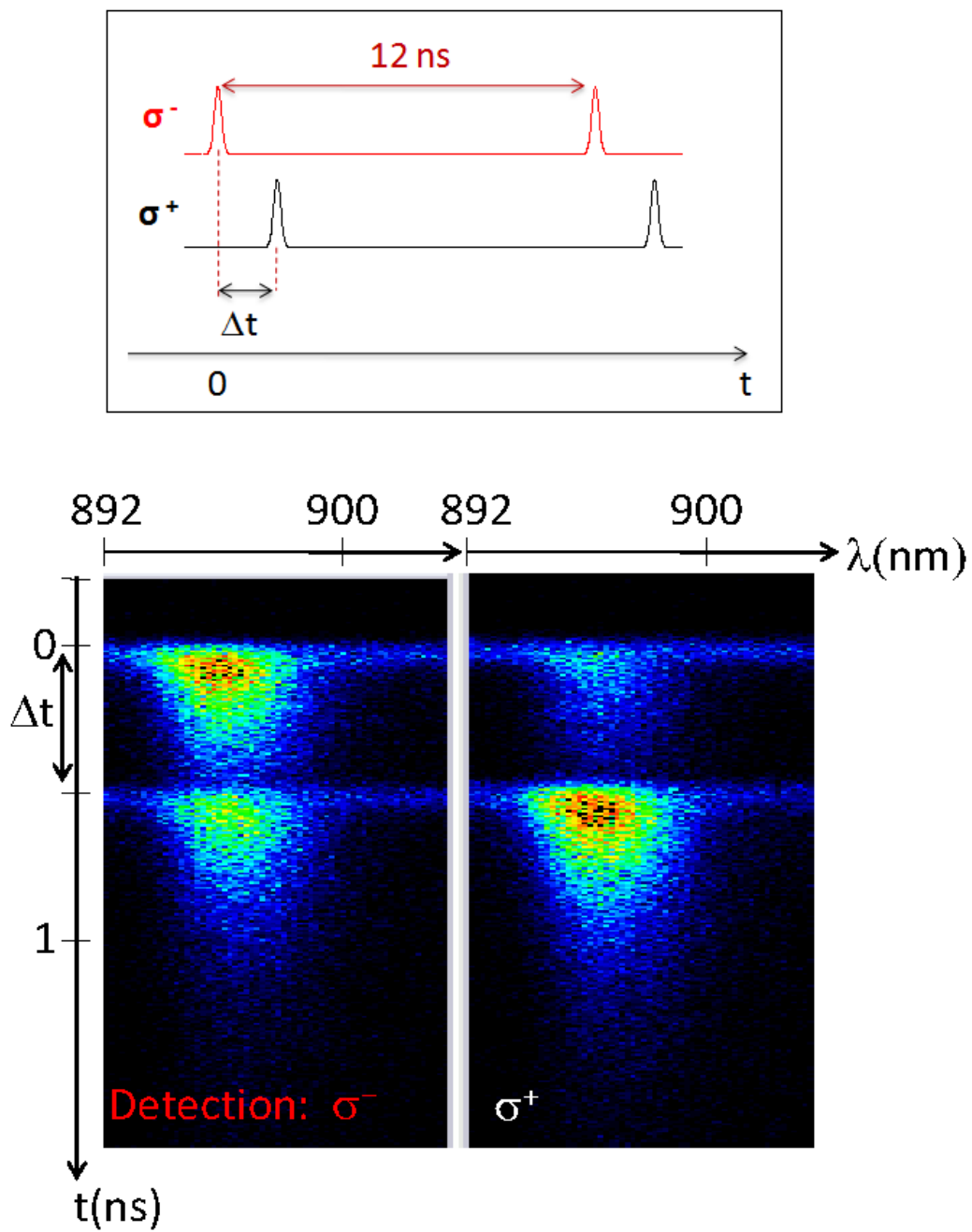

Figure 2. Typical time-resolved PL results from sample MN1 using two excitation beams with opposite circular-polarizations. The time delay between the pulses from the two distinct beam is Δt= 0.5 ns, as shown by the schematic representation on top. The streak camera images correspond to the σ+ and σ- components of the PL emission.

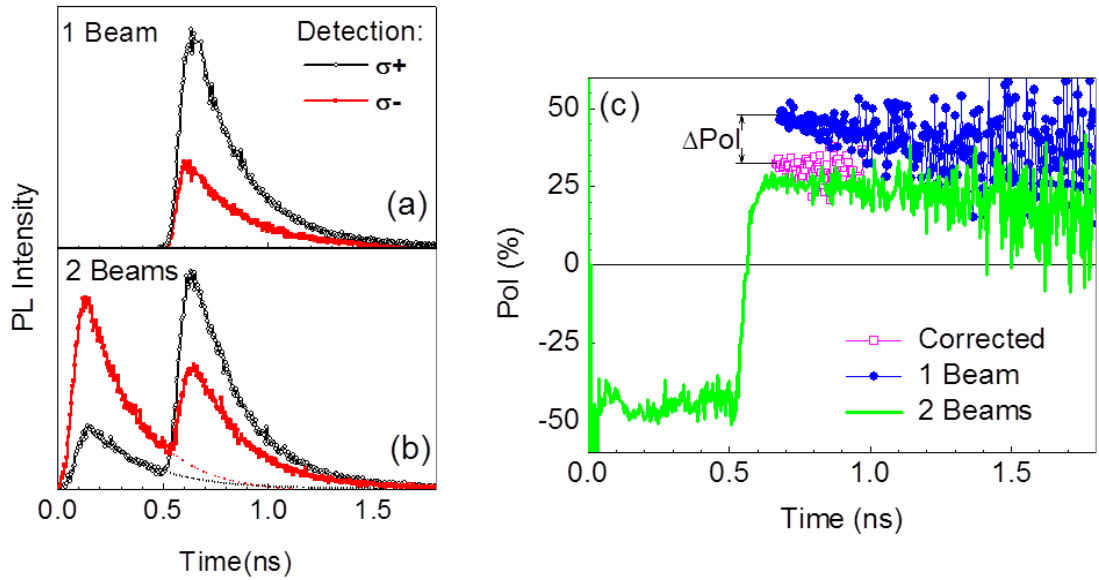

Figure 3. (Color Online) σ⁺ (symbols) and σ⁻ (solid line) PL transients of the QW emission from sample MN1. (a) Under one excitation beam. (b) Under two excitation beams with opposite polarization and a time delay of Δt= 0.5 ns, where the dashed line indicates the expected PL intensity without the second beam. (c) Circular polarization degree for the measurements using two excitation beams (solid green line), one excitation beam (solid blue circles) and the corrected polarization obtained by subtracting the PL intensity from the pre-pulse (open magenta squares).

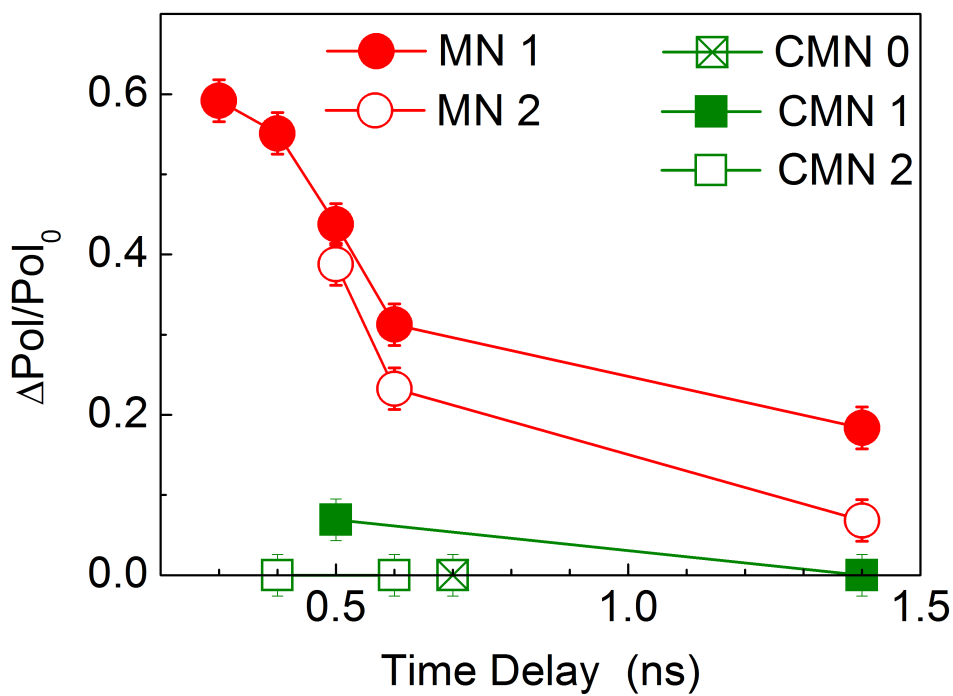

Figure 4. Time-delay (Δt) dependence of the ratio (*ΔPol/Pol₀*) for all the investigated samples.

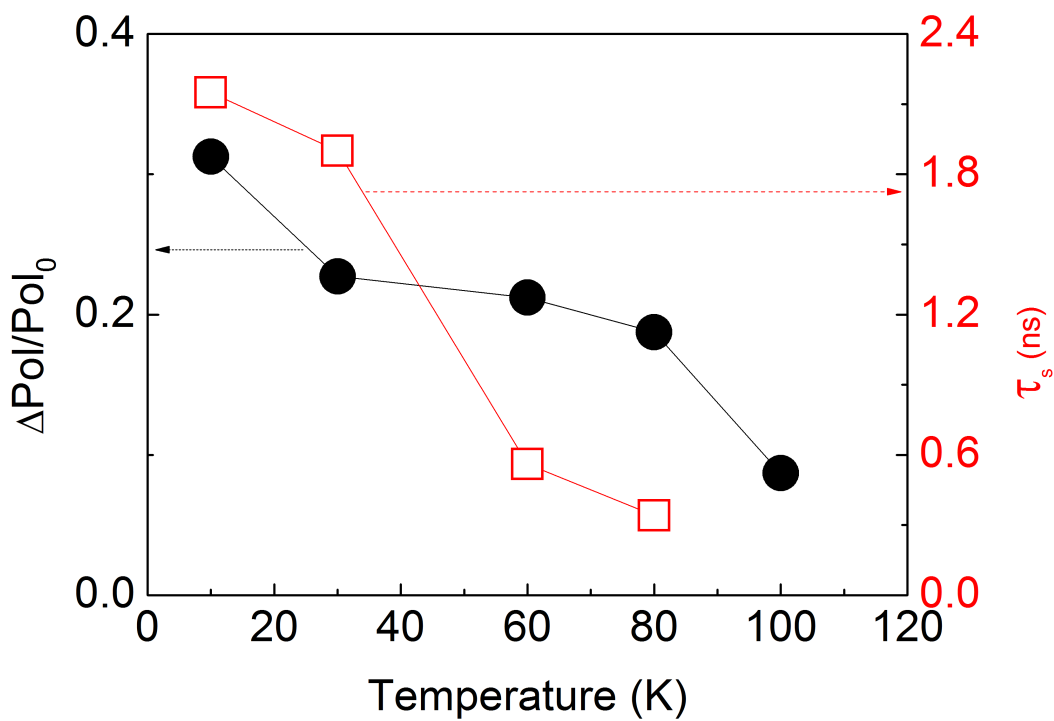

Figure 5. Temperature dependence of the spin relaxation time ($\tau_s$) (open squares) and the ratio (*ΔPol/Pol₀*) obtained using Δt= 0.6 ns (solid circles) for sample MN1.